\documentclass[aps,pre,superscriptaddress,reprint,showpacs]{revtex4-1}
\usepackage{amsfonts}
\usepackage{amsmath}
\usepackage{amssymb}
\usepackage{graphicx}
\usepackage{bm}
\usepackage{xcolor}%
\usepackage{psfrag}

\providecommand{\U}[1]{\protect\rule{.1in}{.1in}}

\begin{document}
\title{Co-evolution of networks and quantum dynamics: a generalization
  of preferential attachment}

\author{Vincenzo Nicosia}
\affiliation{School of Mathematical Sciences, Queen Mary, University
  of London, Mile End Road, London E1 4NS, UK}
\affiliation{Laboratorio sui Sistemi Complessi, Scuola Superiore di Catania, 
  Catania, Italy.}

\author{Takuya Machida}
\affiliation{Japan Society for the Promotion of Science, Japan}
\affiliation{Department of Mathematics, University of California,
  Berkeley, USA}

\author{Richard Wilson}
\affiliation{Department of Computer Science, University of York,
  Deramore Lane, Heslington, York, YO10 5GH, UK }

\author{Edwin Hancock}

\affiliation{Department of Computer Science, University of York,
  Deramore Lane, Heslington, York, YO10 5GH, UK }

\author{Norio Konno}

\affiliation{Department of Applied Mathematics, Faculty of
  Engineering, Yokohama National University 79-5 Tokiwadai, Hodogaya,
  Yokohama, 240-8501, Japan}

\author{Vito Latora}
\affiliation{School of Mathematical Sciences, Queen Mary, University
  of London, Mile End Road, London E1 4NS, UK}
\affiliation{Dipartimento di Fisica e Astronomia, Universit\`a di 
Catania and INFN, 95123 Catania, Italy}  
\affiliation{Laboratorio sui Sistemi Complessi, Scuola Superiore di Catania, 
Catania, Italy.}

\author{Simone Severini}
\affiliation{Department of Computer Science, and Department of
  Physics \& Astronomy, University College London, Gower Street,
  London WC1E 6BT, UK}

\begin{abstract}
We propose a model of network growth in which the network is
co-evolving together with the dynamics of a quantum mechanical system,
namely a quantum walk taking place over the network.  The model
naturally generalizes the Barab\'{a}si-Albert model of preferential
attachment and it has a rich set of tunable parameters, such as the
initial conditions of the dynamics or the interaction of the system
with its environment. We show that the model produces networks with
two-modal power-law degree distributions, super-hubs, finite
clustering coefficient, small-world behaviour and non-trivial
degree-degree correlations.
\end{abstract}

\maketitle

Models of graph growth are important for studying and simulating the
behaviour of a large variety of real-world phenomena and for the
\emph{in silico} construction of networks with given
properties. Growth processes occur ubiquitously in social systems,
technology, and nature. See, for example, Ref.~\cite{new, dor, boc}
for overviews on complex networks growth. Among the most extensively
studied models of network growth are those based on \emph{preferential
  attachment} -- or \textquotedblleft the rich get
richer\textquotedblright scheme -- together with its many
variations~\cite{barabasi}. Predating networks science, the basic idea
behind preferential attachment goes back to the 1920s and the work of
the statistician Yule~\cite{yu}. The set-up usually consists of two
ingredients: an iterative process in which new nodes are sequentially
added to an existing graph; and a mechanism for choosing the
neighbours of newly arrived nodes. Only when the preference on the
neighbours is a linear function of the degrees of the nodes, then the
degree distribution of the growing graph turns out to be a power-law,
as those observed in the majority of real-world complex networks. The
literature contains many variants of preferential attachment,
respectively defined by local rules, fitness, redirection, copying,
substructures, games, geometry, \emph{etc.  }~\cite{vaz}.

We are interested in generalizing the classical preferential
attachment model, in which the choice of the nodes to link to is
specified by the dynamics of random walks~\cite{af}.  The original
idea, illustrated in Refs.~\cite{saramaki, ikeda, evans} is
simple. Imagine a walker moving along the edges of a graph. At a given
node, the walker chooses to stay where it is or to move to one of the
neighbouring nodes with a fixed but arbitrary probability. If we wait
long enough, the probability that the walker is at a specific node is
proportional to the number of its neighbours:\ this probability
converges towards a unique stationary distribution and is independent
of the starting node -- a fundamental property in algorithmic
applications of Markov chains~\cite{mo, sj}. Once we are close enough
to the stationary distribution, we add a new node to the graph, and
choose its neighbours according to the occupation probability
distribution induced by the walker. If we keep adding nodes in this
way, we eventually grow a graph whose degree distribution follows a
power-law. When the walker makes an unbiased choice at each node,
i.e. when the stationary occupation probability of the walker at node
$i$ is a linear function of the degree of $i$, then this mechanism
produces exactly the \emph{Barab\'{a}si-Albert} (BA) \emph{random
  graph} \cite{barabasi}, which is the most widely studied outcome of
preferential attachment so far.

Graph growth indeed fits into a larger picture, namely the study of
dynamical graphs,\emph{ i.e.}, graphs changing in time. This is a
direction that is currently generating interest as a natural
development of static network theory (see Ref.~\cite{sara12} is a
recent review). Letting the structure of a graph \emph{co-evolve}
together with a dynamical process is a particularly appealing and
well-motivated idea.  In particular, the existence or activity of a
node and the strength of a link can be time-dependent on the state of
a dynamical process taking place on the graph. For instance, if the
nodes of a graph represent individuals having an opinion which changes
over time according to a certain rule, and the links stand for
friendship among nodes, then the existence and strength of each link
can change over time as a function of the difference of opinion
between adjacent nodes. People usually tend to remain linked with
neighbours who share similar opinions, and to sever links to other
individuals having different opinions. In this case the structure of
the network depends on the distribution of opinions and, on the other
hand, the opinion dynamics depends on the actual connection
pattern. In a single word, the network and the opinion formation
process are \emph{co-evolving}. Opinion formation is of course only a
very specific example of a process able to drive the evolution of a
network. Many other models of networks co-evolving with
synchronization~\cite{Assenza2011}, diffusion~\cite{Aoki2012}, and
voter models~\cite{Vazquez2008,Bohme2012} have been discussed in the
last decade (see also Ref.~\cite{Gross2008}).

Networks seen as states of a quantum mechanical system co-evolving
together with a classical process have been proposed to explore the
role of emergence in approaches to discrete quantum
gravity~\cite{Caravelli2010}. Networks whose edges correspond to
bipartite states -- essentially certain circuits with two-qubit gates
-- have been studied in relation to entanglement
distribution~\cite{Acin2007}.

With the aim of designing a new methodology to grow complex networks,
we ask the following question:\ what happens when we consider a graph
whose growth depends on the state of a quantum mechanical system? In
particular, we are interested in replacing the random walk dynamics,
which produces the BA random graph, with a quantum dynamics. At each
step of the growth process, the neighbours of the newly added node are
chosen by observing the state of the quantum system -- by using a
standard (von Neumann)\ measurement. We have the co-evolution of two
processes:\ each step of the graph growth process depends on a quantum
dynamics and, conversely, the dynamics takes place in a phase space
(the graph) modified during the growth.

Quantum walks have been extensively studied in the last forty years: a
continuous version is discussed in~\cite{fa} and dates back to 1964; a
discrete version was introduced in the 1990's~\cite{ah}. During the
past decade, quantum walks have acquired an important role in the
context of quantum computation as a methodology for designing
algorithms~\cite{am, ko}.  Quantum walks are also essential in
modeling \emph{quantum buses} for the transfer of information in
nanodevices. The dynamics of an exciton in a spin chain or an
arbitrarily coupled spin system is modeled by a coherent quantum
walk~\cite{bo0, c}. Additionally, the transport of energy in large
molecular complexes has been explained using a class of quantum walks
whose evolution is assisted by interactions of the system with a noisy
environment~\cite{re08}. Very recently, quantum walks have also been
employed for the characterisation of complex
networks~\cite{mulken2011,Sanchez2012}.

The quantum walker, like the classical (\emph{i.e.}, random) one,
induces a probability distribution on the nodes of a graph. However,
the distribution is obtained by measuring the state of the system at a
given time. In quantum mechanics, the state (of a closed system) is
identified with a unit vector in a complex phase space, with
probabilities substituted by amplitudes. A major property of quantum
walks is the existence of interference effects during the dynamics.
Once the position of the quantum walker is measured, the corresponding
probability distribution is the result of an interference
pattern. Notably the distribution does not converge in time because
the evolution of a quantum mechanical system is completely
reversible~\cite{ah, dir}. The dynamics can be periodic, or
quasi-periodic, but there is no convergence unless we take a
time-average~\cite{ah1} -- or we stop the evolution at a given
time. Interference is one of the ingredients that permits algorithms
of good performance to be designed~\cite{aa} and is also responsible
of many counterintuitive behaviours of quantum systems.  For instance,
transporting a packet of information\ from one node to another node
without error, even if routed \textquotedblleft
randomly\textquotedblright~through the graph -- a phenomenon called
perfect state transfer \cite{c}; or reaching far away nodes with an
average probability that is exponentially higher than that for the
classical analogue \cite{cf}. It is also remarkable that quantum walks
have been successfully implemented through various experimental
schemes involving light or matter~\cite{bl, ma, sh}.

Here we work with \emph{continuous-time quantum walks}
(CTQW)~\cite{fg}. In these processes the matrix defining network links
is interpreted as the system Hamiltonian:\ this is the operator
corresponding to the total energy of the system. The Hamiltonian
specifies the interactions between the particles associated with the
nodes in terms of coupling strengths. CTQWs are reversible by
definition, since the dynamics is governed by the Schr\"{o}dinger
equation. (A formal definition is given in Appendix.)

What can we say about graphs whose growth depends on the state of
quantum walks? Do they have structural properties comparable to those
of the BA random graph? The time at which the measurement is performed
(\emph{i.e.} the time at which the walk is stopped) drastically
influences the distribution of attachment probabilities; moreover, the
choice of the node from which the walk is started may affect the
occupation probability distribution measured at a certain time. These
facts reflect the much richer behaviour of quantum walks with respect
to random walks. In order to associate a probability distribution to
the CTQW, we consider the time-average of all distributions obtained
by running the walk for a given time. At each step of the iteration,
we could choose the distribution obtained by stopping the walk at a
time determined by a function of the number of nodes.  We choose here
(arbitrarily) to run the walk for an infinite time -- this is like
stopping the walk at a random time and it looks like a good choice for
exploring the general idea. The time length of each walk is in fact a
tunable parameter, differently from the classical BA model in which
the growth depends on the long-time dynamics of the walk.

\begin{figure*}[!ht]
\centering \includegraphics[width=3in]{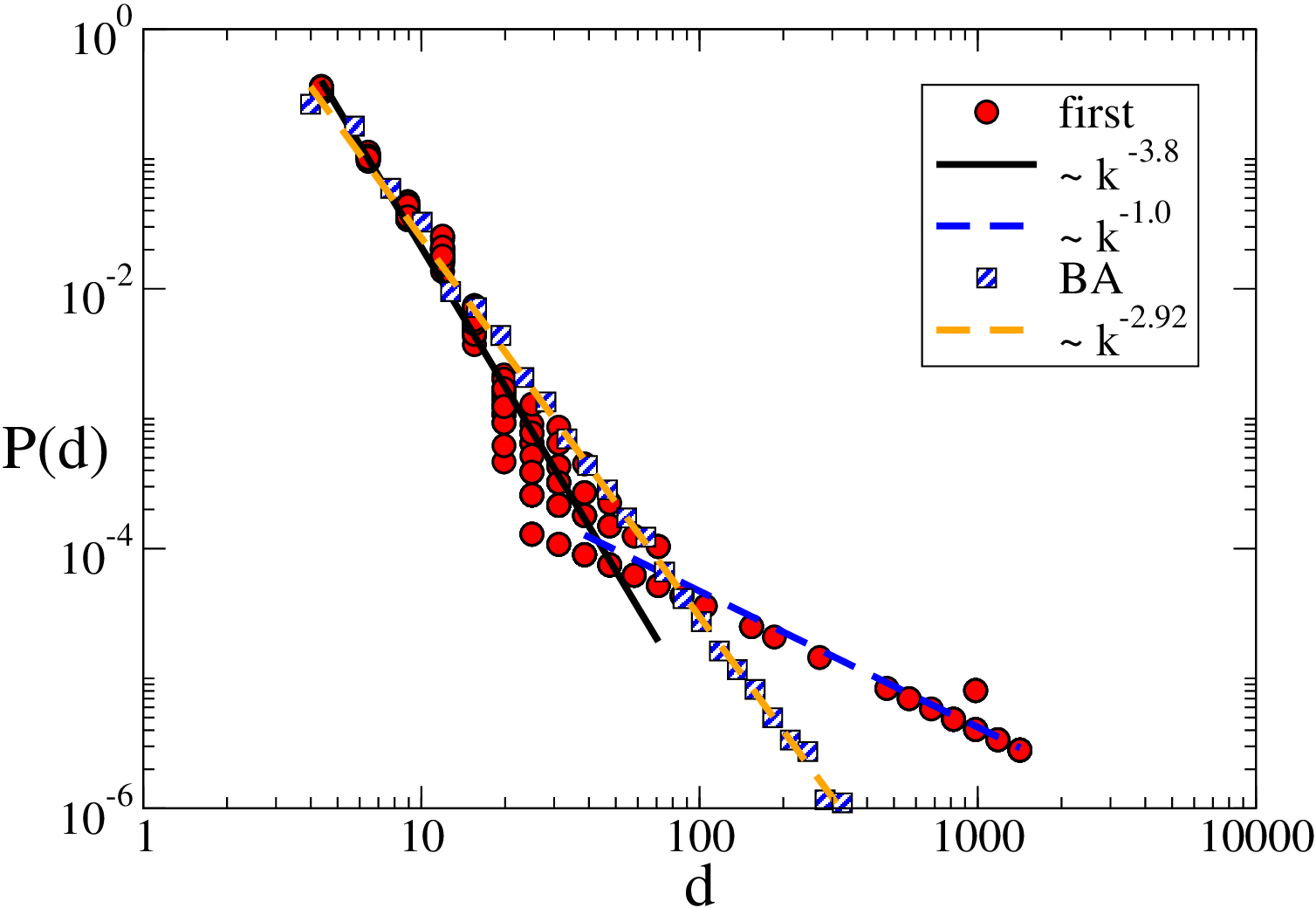}
\includegraphics[width=3in]{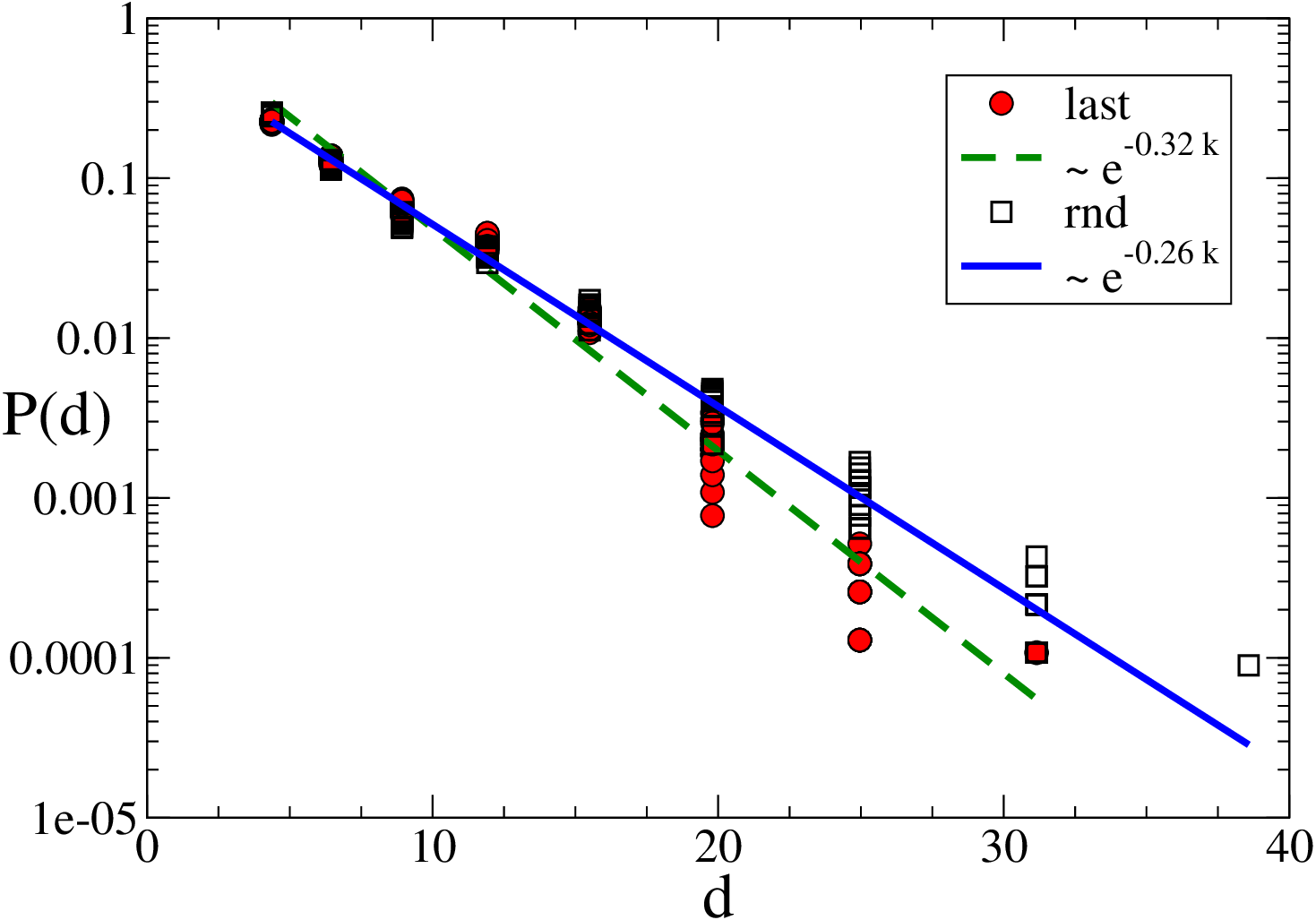}
\\ \includegraphics[width=3in]{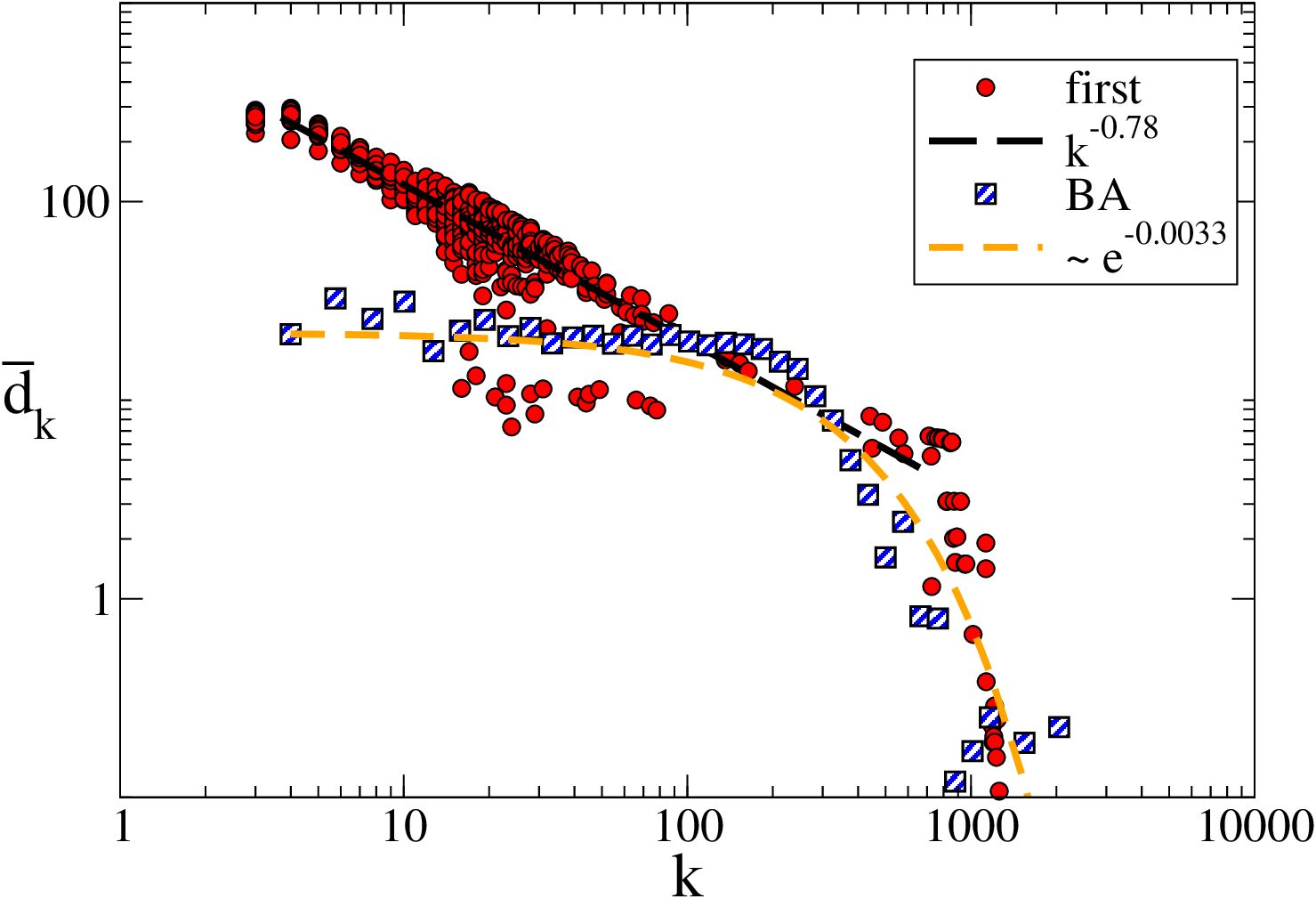}
\includegraphics[width=3in]{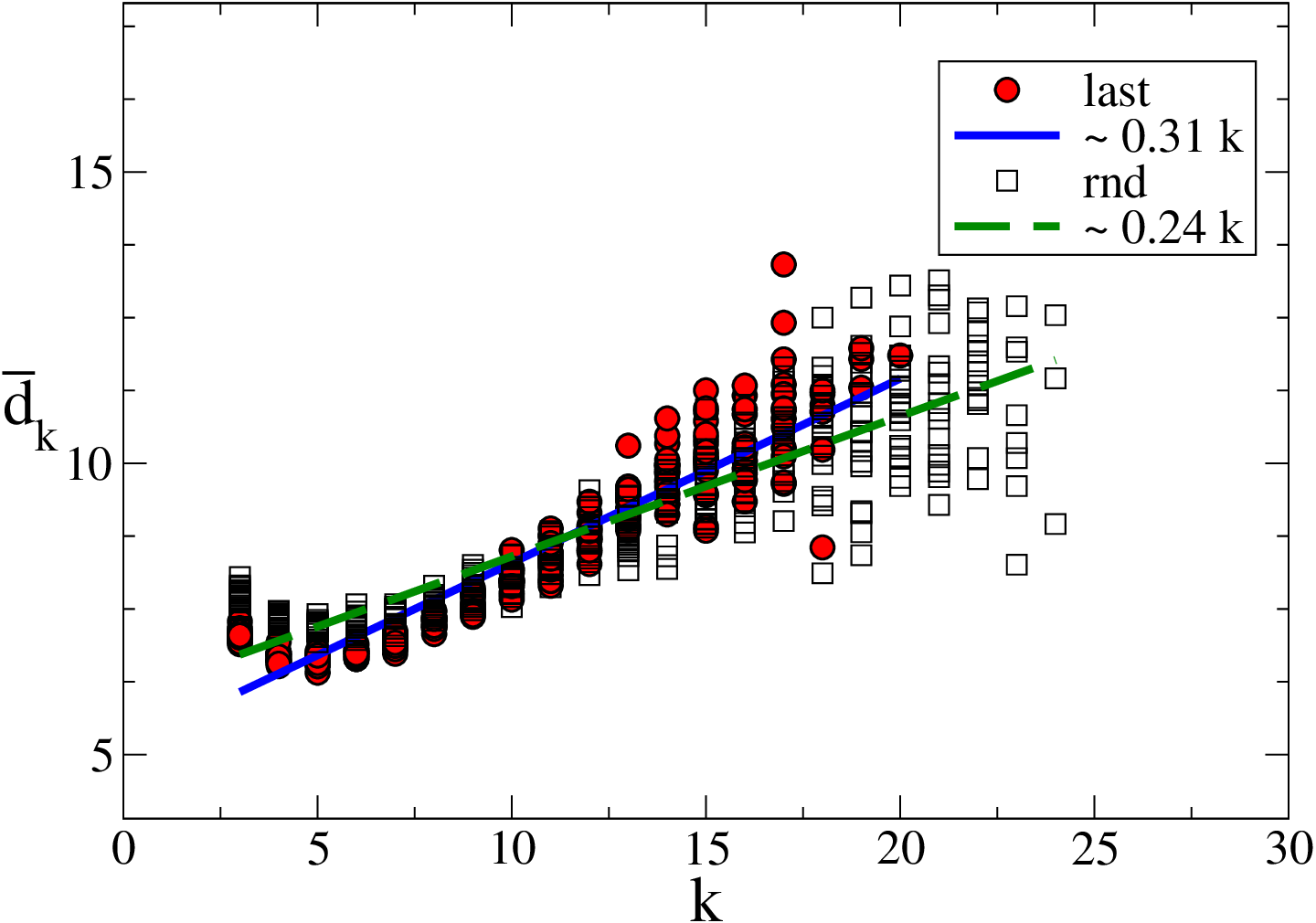} \caption{The
  structural properties of the final network heavily depend on the
  choice of the starting node for the quantum walks. If the walks
  starts from the node used to seed the growth process (upper left
  panel), we get a two-mode power-law degree distribution. In this
  case there is a non-negligible probability of forming a super-hub
  which condenses a large fraction of the edges of the network. Also, the final network 
  has pronounced disassortative
  degree-degree correlations (lower left panel). Conversely, if the
  walk starts from the lastly added node, or from a randomly selected
  one, then the degree distribution of the final network is instead
  exponential (upper right panel). In this case there are slightly
  assortative degree-degree correlations (lower right panel). The
  results are based on 20 realizations with $N=3000$ nodes and
  $K=9000$ edges for each scenario.}%
\label{fig:distr}%
\end{figure*}

We consider here three alternatives for the starting node used to seed
the quantum walk: \emph{a)} the walks always start from the initial
node, \emph{i.e.} the node added at the first step of the graph
growing iteration; \emph{b)} the walks always start from the node
added at the last step; \emph{c)} at each step, the starting node is
chosen at random -- of course, we could consider any probability
distribution on the set of existing nodes. Since the growth is driven
by the quantum walk which effectively acts as a \textquotedblleft
controller\textquotedblright, we expect different asymptotic
distributions. On the one hand, obtaining the time-average of a CTQW
is a tractable problem; but on the other hand, predicting the
properties of our growing graphs is a difficult one because the
time-average has erratic behaviour~\cite{god}. Presenting an analytic
treatments of the asymptotics remains an open problem.

In Fig.~\ref{fig:distr} we report the results of numerical simulations
of the model with different initial conditions. We have constructed
networks with $N=3000$ nodes and $K=9000$ links. We observe that
different choices of the starting node produce graphs with different
structural properties. For selection \emph{(a)}, the final graph is
characterized by a two-mode power-law degree distribution (upper left
panel) and has \emph{super-hubs}, \emph{i.e.}  nodes with degree of
the same order of the total number of nodes. Such exceptionally
highly-connected nodes are usually among the oldest ones, \emph{i.e.}
the nodes added in the very first steps of the iteration. Each
super-hub turns out to be incident with up to $30\%$ of the total
number of edges in the final graph. This condensation phenomenon is
indeed observed in real communication and information networks,
including the Internet and the World Wide Web, and in biological
networks~\cite{boc}. Notice that the degree distribution obtained in
{\em (a)} is different from that obtained for the BA random graph
shown in the same panel. An obvious by-product of the existence of a
super-hub is that the average length of the shortest paths between the
nodes is much smaller than the one observed in the Erd\H{o}s-Renyi
(ER) and BA random graphs of the same size, as highlighted below.

\begin{figure*}[!ht]
  \psfrag{APL}{$\ell$}
  \includegraphics[width=3in]{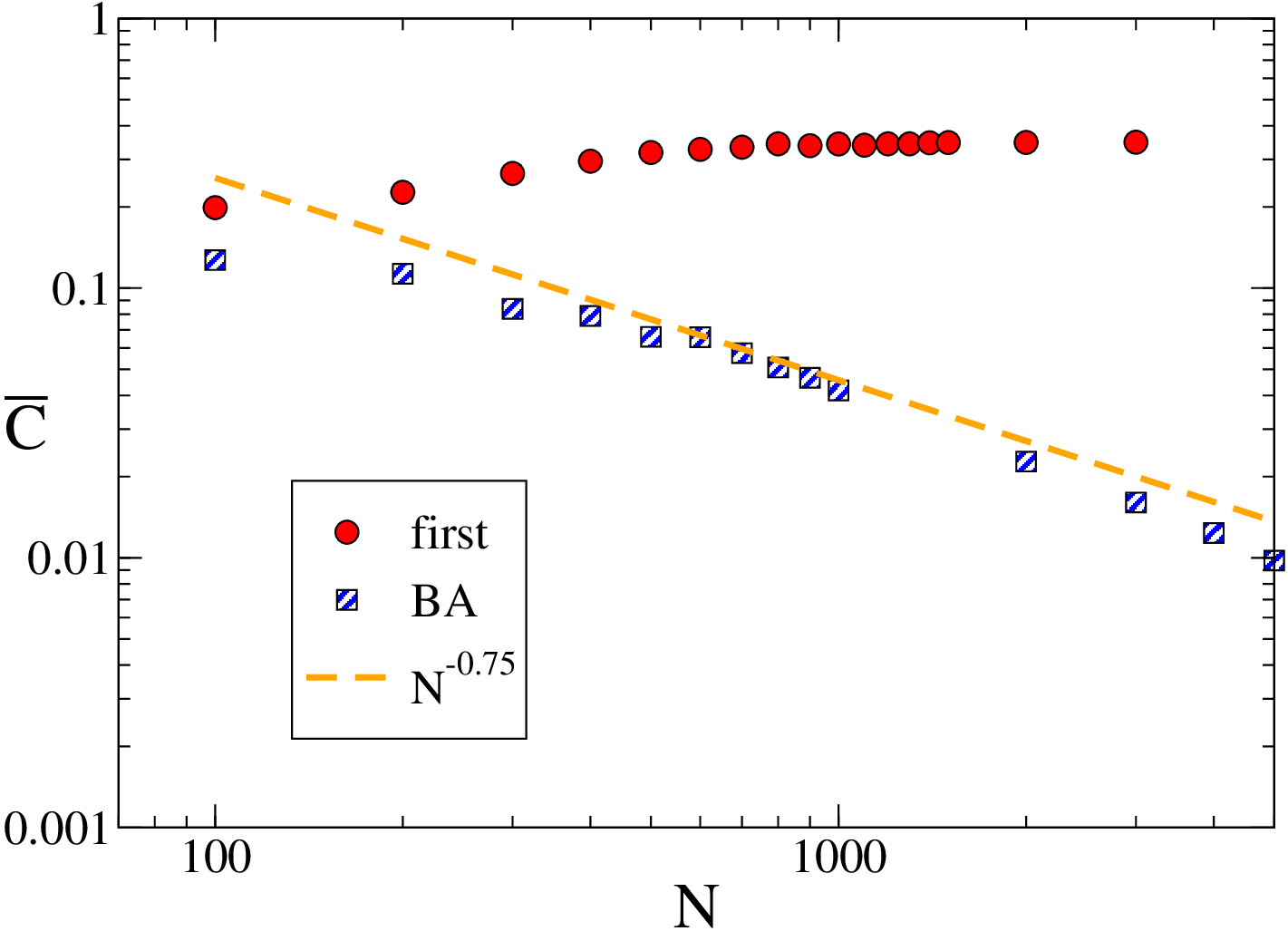}
  \includegraphics[width=3in]{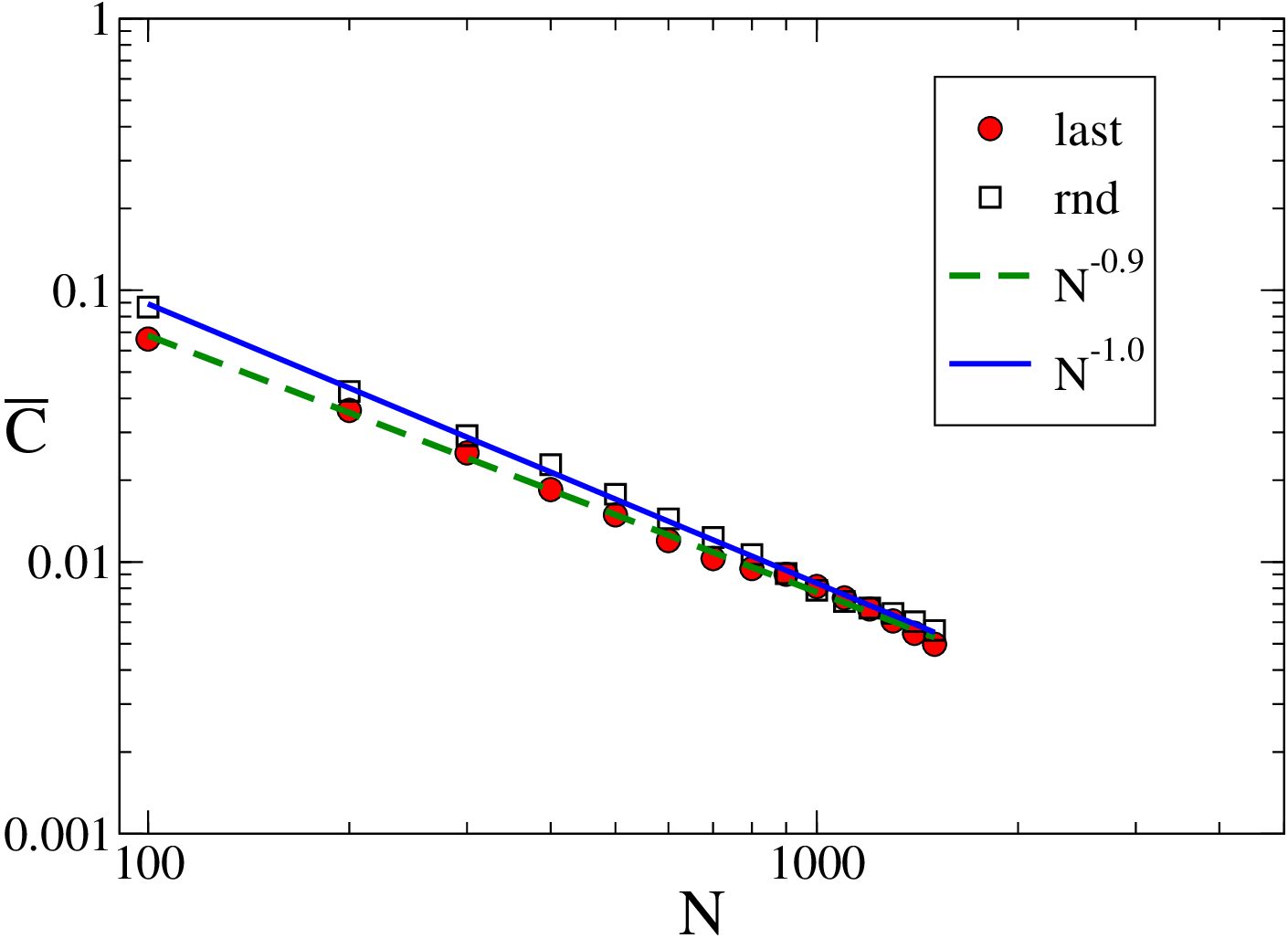}\\
  \includegraphics[width=3in]{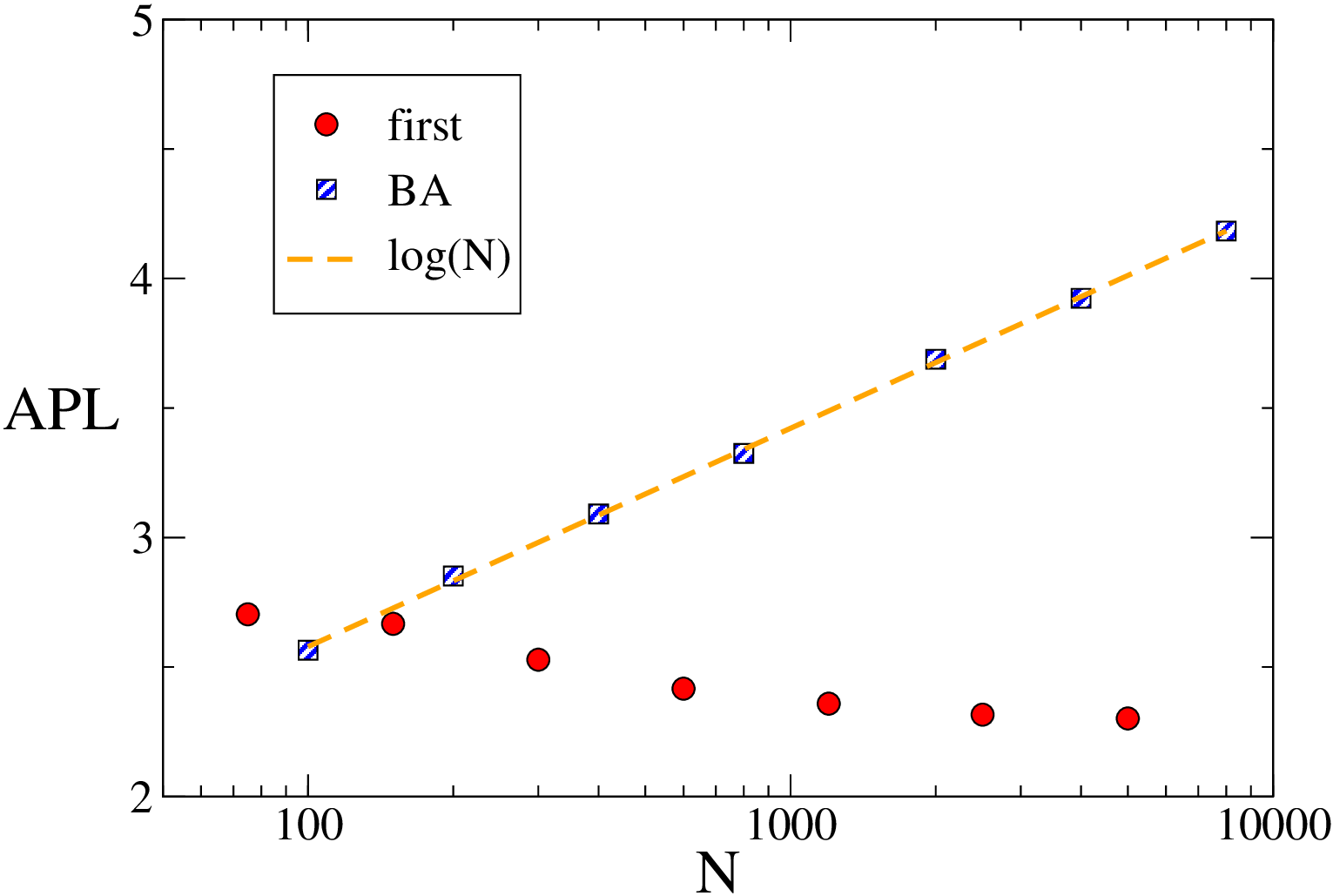}
  \includegraphics[width=3in]{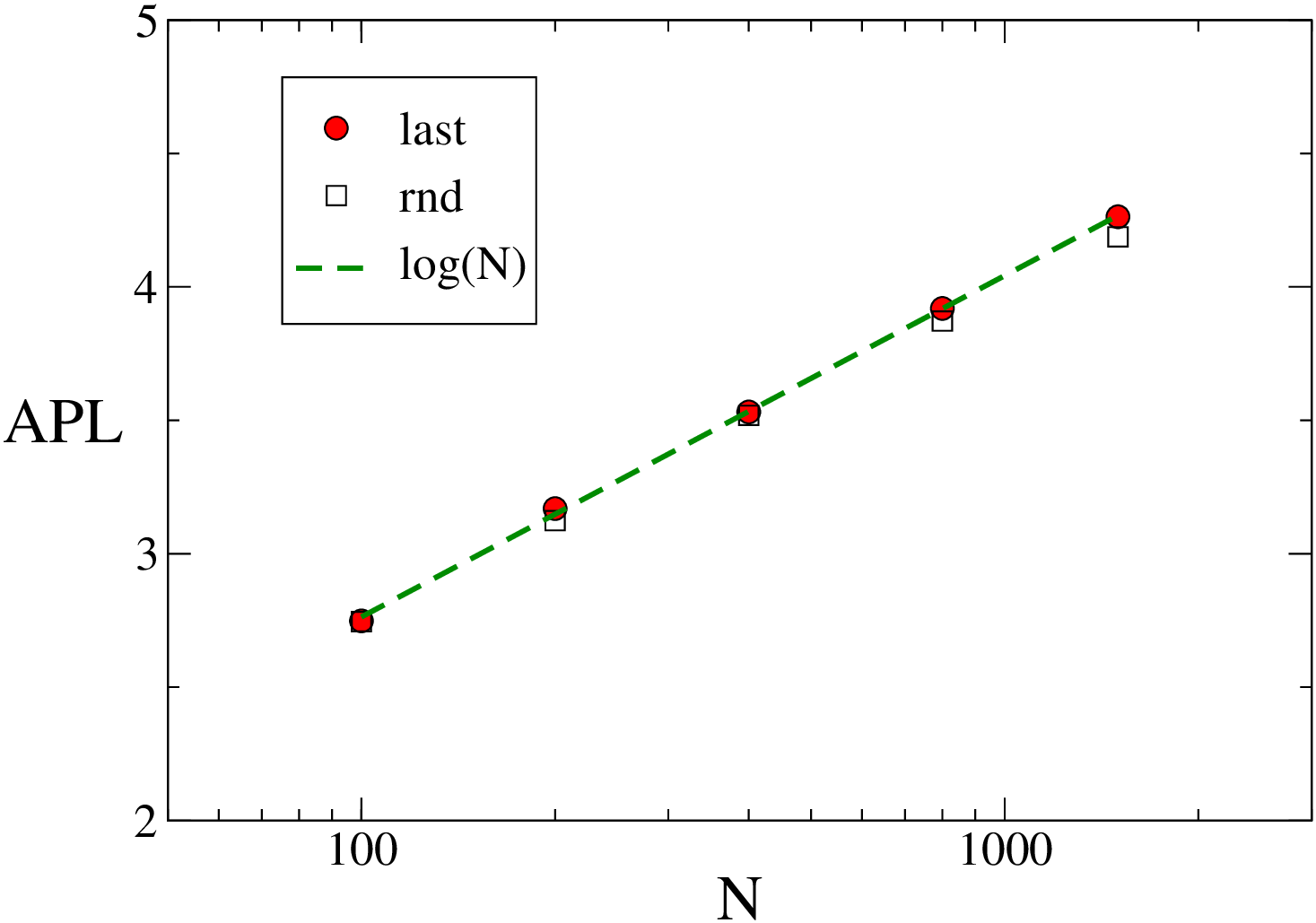}\\
  \caption{Scaling of average clustering coefficient (top) and
    characteristic path length (bottom) for CTQW-based growths. When
    walks start from the first node (left panels) the average
    clustering coefficient increases with the network order $N$ until
    it reaches a plateau around $\overline{C}\simeq 0.34$, while the
    characteristic path length decreases with $N$. If the walk starts
    from the last node or from a randomly selected one, then the
    clustering coefficient decreases over time as $N^{-1}$ while the
    characteristic path length increases logarithmically with $N$.}
  \label{fig:clust_apl}
\end{figure*}

The final graph also exhibits a surprisingly high local cohesion. This
corresponds to a relatively high value of the average clustering
coefficient (denoted by $\overline{C}$), a property which is
extensively found in real networks but is rarely reproduced in models
of random graphs without introducing artificial ingredients. Another
remarkable property is the presence of pronounced disassortative
degree-degree correlations: the average degree $\overline {d}_{k}$ of
the neighbours of a node with degree $k$ depends on $k$ and decreases
as a power-law, $\overline{d}_{k}\sim k^{\nu}$, with $\nu\simeq -0.8$.
The existence of disassortative degree-degree correlations is
partially due to the fact that the degree distribution of the
resulting graphs do not have a structural cut-off~\cite{bpv}, so that
the average degree of the neighbours for a substantial fraction of the
nodes (\emph{i.e.}, those nodes which share an edge with the
super-hub), is dominated by the degree of the super-hub. For
comparison, we report in the same panel the value of
$\overline{d}_{k}$ for the BA random graph, which is practically
independent of $k$, except for the boundary effects observed for high
values of $k$.

Conversely, for selections of type \emph{(b)} and \emph{(c)}, namely when the walk starts either
from a randomly chosen node or from the last node, we obtain
exponential degree distributions and small assortative degree-degree
correlations (upper right and lower right panels, respectively). In
this case, the degrees appear to be more homogeneous: the final graph
has neither hubs nor super-hubs, it exhibits a negligible clustering
coefficient and an average shortest path length (denoted by $\ell$)
comparable to that of an ER or a BA random graph with an equal number
of nodes.

For instance, in ER and BA random graphs with $N=3000$ nodes, as those
reported in Fig.~\ref{fig:distr}, $\ell\simeq 4.29$,
$\overline{C}\simeq 0.0035$ and $\ell\simeq 2.54$, $\overline{C}\simeq
0.02$, respectively. Instead, in the case of CTQW-based growth, when
the walks start from the seed node of the growth process, then the
final graph exhibits a clustering coefficient $\overline{C}\simeq
0.331$, which is much higher than expected in a random graph, and a
considerably smaller average path length $\ell\simeq 2.41$, which is
in turn smaller than those observed in ER and BA graphs of the same
size and order. Therefore, graphs grown with this method are
small--worlds~\cite{Watts1998}. When the walks start from the last
node, we have $\ell\simeq 4.26$ and $\overline{C}\simeq 0.0051$, which
are values comparable with those observed in ER and BA
graphs. Choosing the starting node at random does not seem to result
in a significant difference:\ $\ell\simeq 4.18$ and
$\overline{C}\simeq 0.006$.  In Fig.~\ref{fig:clust_apl} we study the
scaling of $\overline{C}$ and of the average shortest path lenght
$\ell$ with the network order $N$. We compare CTQW-based networks with
BA scale-free graphs.  Notice that, while the average clustering
coefficient of a BA network decreases as
$N^{-3/4}$~\cite{Barabasi2002rev}, in CTQW-based growth with walks
starting from the seed node $\overline{C}$ grows with $N$ until it
reaches a pleateau around $\overline{C}\simeq 0.34$. Similarly, the
average path length decreases with $N$, until it reaches a pleateau
around $\ell\simeq 2.3$. These scaling behaviours are mainly due to
the condensation of edges around super-hubs, which favours the
creation of triangles and contribute to lower the average path
length. As we see from the two right panels of
Fig.~\ref{fig:clust_apl}, if the walks are started from the last node
or from a randomly selected one, then the scalings of $\overline{C}$
and $\ell$ are comparable with those observed in classical random
graphs. The notions of assortative/disassortative degree-degree
correlations, clustering coefficient and average shortest path-length
are standard in the toolbox of network theory. We recall these
definitions in the Appendix.

As mentioned above, CTQWs have reversible dynamics and the (von
Neumann) entropy of any state during the evolution is zero. The
dynamics changes if we include an interaction between the system and
its environment. This introduces \emph{decoherence}, a phenomenon
responsible for the quantum-to-classical transition. Due to
decoherence effects, the system becomes thermodynamically
irreversible. There are various ways to model decoherence in quantum
walks, for example, by monitoring the evolution of the system at a
certain rate. The non-zero probability of performing measurements can
be interpreted as a weak coupling between the quantum system and a
Markovian environment (see Ref.~\cite{k} for a detailed survey of the
topic). Generally, when we increase the decoherence rate, the quantum
features disappear and after a critical point the behaviour of the
system becomes classical. Thus, in the case of a fully decohered
quantum walk, we are able to recover the familiar preferential
attachment induced by classical random walks~\cite{barabasi, saramaki}
-- a random walk can be also obtained algorithmically from a
CTQW~\cite{Childs2007}.  We can interpolate between these two modes by
turning the level of decoherence up or down~\cite{Brun2003,
  Kendon2004}.  For very high decoherence rates, the system will tend
to remain in the initial state due to the quantum Zeno
effect~\cite{Alagic2005}. This phenomenon can be arguably used to
influence the behaviour of the degree sequence by choosing the node
from which starting each walk.

\bigskip

The purpose of the first papers on quantum walks, written in the
context of quantum computing, was to determine whether these dynamics
could be used in algorithmic applications. Indeed, since then, quantum
walks have been very important as a tool to design new quantum
algorithms~\cite{aa,ah1,am}, in both the continuous and the discrete
setting. Two reasons are behind this fact: quantum walks permit, in
some cases, faster convergence towards the limiting distribution than
their classical analogues; the limiting distribution -- in the ways
that this is defined to avoid the lack of convergence due to unitarity
-- often takes the shape of interference patterns that are far away
from uniform.  Interference is responsible for peaks in the
distribution which may turn out to be useful to sample specific
vertices. Therefore, we can use quantum walks to generate exotic
probability distributions with the potential of highlighting certain
subgraphs. The most successful application of this phenomenon is
arguably perfect state transfer, where the state of a single qubit on
a network of quantum spins can be transferred with complete fidelity
between two particles, or periodicity, where after a given time the
whole network is again in a past state. Such a behaviour gives routing
without local control and it can not be found in classical random
walks.  With respect to the topic of the present paper, faster
convergence would imply a more efficient process to simulate network
growth. Of course, the process is quantum and requires a physical
implementation. In addition, quantum walks have many tunable
parameters as we have seen above. By changing, sometimes even
minimally, the initial conditions and transition amplitudes, we can
modify a walk in substantial ways and this freedom in choosing the
parameters allows the growth of networks with different properties,
something which is not classically available unless we steer each
classical walk with extra local rules.

In this paper we have defined a single model and studied some of its
basic features, but investigating the whole potential of quantum walks
in network growth is a completely open direction of research.
Determining the correlation between dynamical parameters and network
properties is a new, unorthodox problem at the interplay of algebraic
graph theory and eigensystem analysis for unitary matrices. Also, we
have not considered Anderson localisation, partial decoherence, and
noisy evolution.  These are all physical features that can be
naturally introducted into models of network growth, and that are well
motivated by the available background. The effect of such features on
networks may, in turn, suggest novel physical insight.  Another
extension of the model could be in the same spirit of
Ref.~\cite{Janson2013}. The probability of attachment is determined by
the outcome of a process external to the network. In our setting, the
process could be a quantum walk on the integers, with the zero on the
number line corresponding to the first vertex of the growth. What type
of networks are grown when sampling from the interference pattern of
the walk?  Again, by the behaviour of the amplitudes and a specific
tuning of the parameters, we should obtain networks without a known
unified classical method of construction. This is left for further
study.

\emph{Acknowledgments. }Part of this work has been done at the Kavli
Royal Society International Scientific Centre during the workshop
\textquotedblleft Function Prediction in Complex
Networks\textquotedblright\ (28-29 May 2012).  We are grateful to
Gorjan Alagic, Andrew Childs, Vivien Kendon, and Andrea Torsello, for
useful discussion.


\appendix

\section*{Appendix}

\bigskip

\noindent\emph{Quantum walks}

\bigskip

A \emph{graph} $G=(V,E)$ is an ordered pair,
where $V(G)$ is a set whose elements are called \emph{nodes} and
$E(G)\subseteq V(G)\times V(G)$ is a set whose elements are called
\emph{edges}. Since we consider graphs growing over time, we denote by
$G_{t}=(V,E)$ the configuration of nodes and edges at iteration $t$. Notice
that the time of the growth is a discrete variable which is increased
by one for each new node added to the graph. Consequently, the graph
$G_{t}$ has exactly $t$ nodes. The \emph{lazy walk matrix }on a graph
at time $t$, $G_{t}=(V,E)$, is (or, equivalently, is induced
by)$\ W\left( G_{t}\right) =\frac{1}{2}(I_{t}+A\left( G_{t}\right)
\Delta\left( G_{t}\right) ^{-1})$, where $I_{t}$ is the $t\times t$
identity matrix, $A\left( G_{t}\right) $ and $\Delta\left(
G_{t}\right) $ are the \emph{adjacency matrix} and the \emph{degree
  matrix} of $G_{t}$, respectively.  Recall that $[A\left(
  G_{t}\right) ]_{i,j}=1$ if $\{v_{i},v_{j}\}\in E\left( G_{t}\right)
$ and $[A\left( G_{t}\right) ]_{i,j}=0$, otherwise; $[\Delta\left(
  G_{t}\right) ]_{i,j}=\delta_{i,j}d\left( v_i\right) $, where $d\left(
v_{i}\right) :=|\{v_{j}:\{v_{i},v_{j}\}\in E\left( G_{t}\right) \}|$
is the \emph{degree} of $v_{i}$, and $\delta_{i,j}$ is the Kronecker
delta. The \emph{rule} of the walk on the $t$-iteration graph is
$W^{s}\left( G_{t}\right)
\overrightarrow{v}_{i}\longmapsto\overrightarrow{\psi}_{s}$, where
$\overrightarrow{v}_{i}$ is an element of the standard basis of
$\mathbb{R}^{t}$ and $\overrightarrow{\psi}_{s}\in\mathbb{R}^{t}$. The
matrix $W^{s}\left( G_{t}\right) $ induces a distribution on the nodes
of $G_{t}$.  The $j$-point of the distribution corresponds to the
probability of finding the walker at node $v_{j}$ at time $s$ if the
walk started at time $0$ from node $v_{i}$. Thus, the probability is
$\mathbb{P}\left[ i\rightarrow j,s\right]
=\langle\overrightarrow{v}_{j},\overrightarrow{\psi}_{s}\rangle$.

Independently of the initial state, the lazy walk $W\left(
G_{t}\right) $ converges to a unique stationary (probability)
distribution $\mathbf{\pi}\left( G_{t}\right) $, such that
$[\mathbf{\pi}\left( G_{t}\right) ]_{i}=d\left( v_i\right) /2\left\vert
E\left( G_{t}\right) \right\vert $, for each $i=1,...,t$ (see,
\emph{e.g.}, \cite{lovasz}). Convergence is guaranteed by the
stochasticity of $W\left( G_{t}\right) $ and by the fact that there is
a non-zero probability for the walker to remain at each node. The rate
of convergence depends on the spectral gap of the adjacency matrix.

The stationary distribution of a lazy walk on $G_{2}$ is clearly the
vector $\mathbf{\pi}\left( G_{2}\right) =\frac{1}{2}\left[ 1,1\right]
^{T}$. When adding $v_{3}$ to $G_{2}$, we define $\mathbb{P}\left[
  \{v_{1},v_{3}\}\in E\left( G_{3}\right) \right] =\mathbb{P}\left[
  \{v_{2},v_{3}\}\in E\left( G_{3}\right) \right] =\frac{1}{2}$, which
follows from $\mathbf{\pi}\left( G_{2}\right) $. More generally, when
adding a node $v_{t+1}$ to $G_{t}$, we attach $v_{t+1}$ to
$m\geq1$\ nodes in $G_{t}$, so that the probability of attaching
$v_{t+1}$ to $v_{i}$ reads $\mathbb{P}[\{v_{t+1},v_{i}\}\in E\left(
  G_{t+1}\right) ]=[\mathbf{\pi}\left( G_{t}\right) ]_{i}$. The
parameter $m$ is fixed but arbitrary. It is important to remark that
$m$ is not necessarily the degree of node $v_{t+1}$ at the end of the
growth process which may occur at a time $T>t$. When $m>1$ we usually
start the growth from a (connected) graph $G_{m}$.

This mechanism constructs exactly the scale-free graphs for the
original version of the Barab\'{a}si-Albert (BA)
model~\cite{barabasi}. In the BA model, bypassing the walk, a node
$v_{t+1}$ of degree $m$ is added at time $t$. The probability that
$v_{t+1}$ is adjacent to $v_{i}$ is in fact
$\mathbb{P}[\{v_{t+1},v_{i}\}\in E\left( G_{t+1}\right) ]=[\mathbf{\pi
  }\left( G_{t}\right) ]_{i}=d\left( v_i\right) /2\left\vert E\left(
G_{t}\right) \right\vert $, which is exactly the stationary probability
of finding a lazy random walker in $G_{t}$ at node $v_{i}$.

By generalizing the above picture, given a graph on $t$ nodes,
$G_{t}$, we define a unitary matrix $U\left( s,t\right) =e^{-iA\left(
  G_{t}\right) s}$, where $s\in\mathbb{R}^{+}$. \emph{Unitary} means
that $ U \left(s,t\right)
U^{\dagger}\left(s,t\right)=U^{\dagger}\left( s,t\right) U\left(
s,t\right)=I$, where $I$ is the identity matrix and
$U^{\dagger}\left(s,t\right) $ is the adjoint of $U\left(
s,t\right)$. In this case the dynamics is reversible / non-dissipative
because of unitarity. The matrix $U\left( s,t\right)$
defines a \emph{continuous time quantum walk} (CTQW) on $G_{t}$
\cite{am}. The \emph{rule} of the CTQW at the $t$-iteration is
$U\left( s,t\right) |v_{i}\rangle\longmapsto|\psi _{s}\rangle$, where
$|v_{i}\rangle$ is an element of the standard basis of a formal
Hilbert space $\mathcal{H}\cong\mathbb{C}^{t}$ and $|\psi_{s}\rangle
\in\mathcal{H}$. The Dirac notation tells that $\left\Vert |\psi_{s}%
\rangle\right\Vert =1$. The probability that at time $s$ the walker
visits a node $v_{j}$ starting in a node $v_{i}$ is $\mathbb{P}\left[
  i\rightarrow j,s\right] =\left\vert [U\left( s,t\right)
]_{i,j}\right\vert ^{2}$. This probability is obtained by a projective
measurement on $|\psi_{s}\rangle $:\ $\mathbb{P}\left[ i\rightarrow
  j,s\right] =\left\vert \langle
v_{j}|\psi_{s}\rangle\right\vert^{2}$. The vector (or
ray)$\ |\psi_{s}\rangle$ contains the amplitudes associated to each
element of the standard basis. The measurement transforms amplitudes
into probabilities. According to the axioms of quantum mechanics the
post-measurement state is the observed standard basis vector.

The matrix $A\left(  G_{t}\right)  $ is interpreted as the Hamiltonian
inducing the quantum mechanical evolution of a particle whose degrees of
freedom of the dynamics are constrained on the edges of $G_{t}$. Indeed, this
can be seen as the operator describing the evolution of the single excitation
sector of a quantum spin system (XY model) in virtue of the Jordan-Wigner
transformation \cite{bo0}.

At the $t$-th graph iteration, the \emph{mixing matrix} of the CTQW at time $s$ is
defined by%
$$
  M_{s,t}=e^{iA\left(  G_{t}\right)  s}\circ e^{-iA\left(  G_{t}\right)
    s}=U\left(  s,t\right)  \circ U\left(  -s,t\right)  ,
$$
where $[A\circ B]_{i,j}:=[A]_{i,j}\cdot\lbrack B]_{i,j}$ denotes the
Schur-Hadamard product of two matrices $A$ and $B$. The matrix $M_{s,t}$
depends on $s$; it gives the \emph{instantaneous mixing} behaviour of the
walk. Formally, an element of $M_{s,t}$ is constructed by multiplying together
the amplitudes obtained by evolving the system for a time $s$ in the future
and for a time $s$ in the past. Differently from the case of a lazy random
walk, here there is never convergence, because the dynamics is
non-dissipative. For the walk, we have $\left\Vert |\psi_{s}\rangle\right\Vert
=\left\Vert U\left(  s,t\right)  |v_{i}\rangle\right\Vert =1$.

We could also define an \emph{instantaneous mixing time} by looking at the
smallest $s\in\mathbb{R}^{+}$ for which the probability induced by the CTQW is
close in some measure of similarity (for example, total variation distance) to
the uniform or the stationary probability distribution. A possibly different
growth model can be defined by making use of the distribution obtained at a
given time $s$. In this case, the growth is entirely dependent on the chosen
value of $s$; this could be fixed for each $t$ or as a function of $t$,
for instance. To avoid a dependence on $s$, we consider a time-average of
$M_{s,t}$.

The \emph{average mixing matrix} is defined by taking a Cesaro mean:
$$
  \widehat{M}_{t}=\lim_{s\rightarrow\infty}\frac{1}{s}\int_{0}^{s}e^{iA\left(
    G_{t}\right)  s}\circ e^{-iA\left(  G_{t}\right)  s}ds=\sum_{j}E_{j,t}%
  ^{\circ2},
$$
where $E_{r}$ is the $r$-th idempotent of the spectral decomposition of
$A(G_{t})=\sum_{j}\lambda_{j}E_{j}$. In other words, $E_{j,t}$ represents the
orthogonal projection onto the eigenspace ker$(A(G_{t})-\lambda_{j}I)$, where
$\lambda_{j}$ is the $j$-th eigenvalue of $A(G_{t})$. The $ij$-th entry of
$\widehat{M}_{t}$ is the average probability that a walker is found at node
$v_{j}$ (starting at node $v_{i}$). Remarkably, $\widehat{M}_{t}$ is rational
\cite{god}.

In our model of growth based on CTQWs, the attaching probability is defined by
$\mathbb{P}[\{v_{t+1},v_{j}\}\in E\left(  G_{t+1}\right)  ]=[\widehat{M}%
_{t}]_{i,j}$, if we assume that the walker started from node $v_{i}$ at the
$t$-th iteration of the growth process. Depending on the starting node $v_{i}%
$, we get a different attaching probability which will be completely defined
by $G_{t}$. The time length of the walk is not relevant given that
$\widehat{M}_{t}$ is defined as a limit for $s\rightarrow\infty$.

\bigskip

\noindent\textbf{Algorithm.} 
Let $K_{n}$ denote the \emph{complete graph} on $n$ vertices. This is the
unique graph with $n(n-1)/2$ edges. Let $[A]_{i}$ the $i$-th row of a matrix
$A$. The growth of a graph based on CTQWs starts
with $G_{m}=K_{m}$. Then, for every $t>m$, we sample $m$ neighbours $v_{j_{1}%
},v_{j_{2}},...,v_{j_{m}}$ of the new node $v_{t}$ from the distribution
$[\widehat{M}_{t-1}]_{i}$ and create $m$ edges $\{v_{t},v_{j_{1}}%
\},\{v_{t},v_{j_{2}}\},...,\{v_{t},v_{j_{m}}\}$.

The edges are all added at the same time, after $m$ distinct CTQWs have been
performed on $G_{t-1}$. The starting node $v_{i}$ of the CTQW can be
arbitrarily chosen. In the main body of the paper we report the results 
obtained for three different choices of the starting node, namely: \emph{a)}
the first node, \emph{b)} the last node added to the graph and \emph{c)} a
different randomly sampled node for each step of the algorithm. The initial condition
$G_{m}=K_{m}$ can be also relaxed.

This simple growth algorithm, based on the sampling of new edges according to
the time-average of the attaching probability distribution, suffers from the
fact that the evaluation of the Cesaro mean requires the full spectrum of the
adjacency matrix $A(G_{t})$. This is the critical step. In fact, although
efficient schemes exist to compute the few largest eigenvalues of a symmetric
matrix of size $n$, the time complexity of the computation of the whole
spectrum is $\sim O(n^{3})$. At a first analysis, it follows that the number
of steps needed to sample graph of $t$ nodes constructed with our method is of
the order $O(t^{3})$. Sampling a CTQW-based graph is much more costly than
sampling a BA random graph, for which the most efficient algorithm runs in
$O(t)$.

\bigskip

\bigskip

\noindent\emph{Network metrics}

\bigskip

Let $G=(V,E)$ be a graph on $n$ nodes $\{v_{1},v_{2},...,v_{n}\}$. The
\emph{average degree} of $G$ is
$\overline{d}(G)=\sum_{i=1}^{n}d(v_{i})$. Let $d(v_{i})=k$, for a
given node $v_{i}\in V(G)$; then the \emph{average degree of the
  neighbours} of $v_{i}$ is denoted by $\overline{d}_{k}$. We say that
two nodes $v_{i},v_{j}\in V(G)$ are \emph{connected} if there is
$l\in\mathbb{Z}^{+}$ such that $[A^{l}(G)]_{i,j}>0$. Equivalently,
$v_{i}$ and $v_{j}$ are connected if there is a walk from $v_{i}$ to
$v_{j}$. A \emph{walk from }$v_{i}$\emph{ to} $v_{j}$ is a sequence of
edges $\{\{v_{i}=i_{0}%
,i_{1}\},\{i_{1},i_{2}\},\ldots,\{i_{n-1},i_{n}=v_{j}\}\}$, where the
nodes are not necessarily all distinct. When the nodes of a walk are
all distinct then the we call it a \emph{path}. The \emph{length} of a
path is the number of edges in the path. The \emph{distance} $d(i,j)$
between $v_{i},v_{j}\in V(G)$ is defined as the length of the path
from $v_{i}$ to $v_{j}$ with the minimum number of edges. The
\emph{average shortest path length} of $G$ is then defined as
$\ell:=\frac{1}{n(n-1)}\sum_{i,j}d(i,j)$. If there is no path
containing $v_{i}$ and $v_{j}$ then $d(i,j)=\infty$, by convention.
Consequently, $\ell$ is finite only for \emph{connected graphs},
\emph{i.e.}  when every pair of nodes of the graph is connected by a 
path. The graphs generated by the algorithm are connected by
construction. The \emph{clustering coefficient} $C_{i}$ of a node
$v_{i}\in V(G)$ is a measure of the local cohesion at $v_{i}$. Taking
$k=d(v_{i})$, we have $C_{i}:=\frac{1} {k(k-1)}T(v_{i})$, where
$T(v_{i})$ is the number of different triangles containing $v_{i}$. A
triangle is a graph of the form
$(\{v_{i},v_{j},v_{k}\},\{\{v_{i},v_{j}\},\{v_{j},v_{k}\},\{v_{i},v_{k}\}\})$. The
\emph{clustering coefficient} of $G$ is the average of the clustering
coefficients of all nodes: $\overline{C}=\frac{1}{n}\sum_{i}C_{i}$
(See \cite{Newman2003rev} for a general reference on these notions).

Real networks usually exhibit degree-degree correlations. For
instance, in some networks (mostly social, information and
communication networks) high-degree nodes are preferentially linked to
other high-degree nodes, while in biological and technological
networks high-degree nodes are preferentially linked to low-degree
nodes~\cite{new}. The existence of degree-degree correlations can be
quantified in different ways. One of the most common methods is by
computing $\overline{d}_{k}$ as a function of $k$. If a graph is
uncorrelated then the degree of the neighbours of a node of degree $k$
does not depend on $k$, and it is possible to show that
$\overline{d}_{k}=\langle k^{2}\rangle/\langle k\rangle$. In real
networks we observe that $\overline {d}_{k}$ depends on $k$: if
$\overline{d}_{k}$ increases with $k$, we say that the network has
\emph{assortative degree-degree correlations}; on the contrary, if
$\overline{d}_{k}$ decreases when $k$ increases, we say that the
network has \emph{disassortative degree-degree correlations}.
(See~\cite{Newman2003} for an in-depth discussion about degree
correlations in networks.) In most real networks,
$\overline{d}_{k}\sim k^{\nu}$ (with a little abuse of notation) and
the exponent $\nu$ can be effectively used to quantify degree-degree
correlations: $\nu>0$ and $\nu<0$ define \emph{assortative } and
\emph{disassortative networks}, respectively~\cite{Vazquez2002}. The
larger the modulus of $\nu$, the stronger are the degree-degree
correlations.

\end{document}